\documentstyle[12pt]{article}
\begin{document}

\title{Particle/String field theory induced by W-gravity non-linear sigma
models \\and the\\ Grand String Symmetry }
\author{Harry Gasparakis \\Institute for Theoretical Physics \\SUNY at Stony
Brook, N.Y. 11790  \\e-mail: gasparakis@max.physics.sunysb.edu }
\date{July 25, 1993}
\maketitle

\abstract
Preliminary investigations of the topological phase of string theory along the
lines of a (restricted) $\dot{w}_{\infty}$ non-linear sigma model are provided.
Gauge fixing the w gravity gauge fields by preserving a geometric identity
Lorenz covariantizes the w-particle and gauge covariantizes the YM. The notion
of foliation ghosts is introduced. Connection between $\dot{w}_{\infty}$ and
homotopy associative algebras is indicated.  The second quantized twistor space
of string theory is constructed and compared with the first quantised one.
Speculations about the relevance of the massive string fields to the
compactification of the bosonic string down to four dimensions times the
standard model group as well as to the solution of the background independence
problem and the integration over moduli space problem are also provided.

\section{Introduction/Motivation}

There have been some claims in the literature (\cite{kn:eva} , \cite{kn:gro})
that actually string theory is equipped with bigger symmetry than it is usually
realised, and that this symmetry is broken by the vacuum expectation values of
the spacetime fields.By this symmetry, had it been unbroken, one would be able
to think about the higher mass modes of the string as actually being (massless)
gauge fields. It has also been argued that this symmetry probably cannot be
manifest in a spacetime-realised phase ( \cite{kn:eva}, \cite{kn:giv}) and it
usually goes by the name `topological phase' of string theory. By the results
of \cite{kn:gro} about the existence of linear relationships between the string
scattering amplitudes in the high energy regime, one suspects that this
symmetry is linear\footnote{However note the attempts to construct
$W_{3}$-based string theories e.g. \cite{kn:prs}.}, and and pressumably
$w_{\infty}$ or some quantum extension (\cite{kn:Winf}), but no explicit
investigation was ever (up to m
y knowledge) provided. Also the geometrical interpretation of $w_{\infty}$
(\cite{kn:bak}, \cite{kn:ber} ,\cite{kn:hul1}, \cite{kn:hul2}) as Finsler-type
geometry, or as area preserving diffeomorphisms does not have a clear stringy
interpretation and also it is not clear their relation to the Teichmuller space
of the theory.  A related  problem is the expected equivalence of the BRST
string field theory (\cite{kn:sie}, \cite{kn:witb}) and the sigma model
approach, although the field redefinitions that will make this equivalence
manifest are not known in general (see however \cite{kn:mamu}). Another problem
that we will show is related to the above  is the problem of the acausal first
quantised vacuum of the theory.

 Let me recollect some known facts: One can start with the usual open bosonic
string field theory action, with the usual kinetic term involving the BRST
operator of the first quantised string. One knows that by shifting the
background we can still write the theory in the same form, with a different
BRST operator, and a different product between states (\cite{kn:sen}). It is
conjectured in the literature that this theory with the shifted background
corresponds to a different conformal field theory, if and only if the
background is conformal, however a complete proof involving the explicit field
redefinitions nessesary does not exist exept for special simple cases. What is
even less clear is what happens if the background is not conformal. The precise
known statement (in the open string case for simplicity) is as follows:

The result of perturbing the BRST operator of the original theory by a state A,
where
\[A=\sum_{{r}}A_{{r}}(x)|{r}\rangle,\]

where the summation runs over all states of a conformal  field theory  with
target space dependent coefficients, is known (\cite{kn:sen}) to be
\begin{equation}
Q \longrightarrow Q_{A}=Q+[A,\;] ,\label{eq:2}
\end{equation}
and it is also known that this new theory is consistent if the background is
conformal.
It is intuitively clear that doing this is equivalent to starting with a sigma
model action involving the background fields as they appear in A, so therefore
involving W-gravity background fields  which come into play from the higher
spin states of the conformal field theory, and then building the string field
theory using the BRST operator of this W-gravity sigma model lagrangian. A
natural conjecture then would be that this approach would lead to the solution
of the background independence problem, the background being defined by the
W-gravity currents, provided that the BRST operator we will use in constructing
the string field theory will be the one of the most general w gravity sigma
model, including the two dimensional dilaton, and also the w-ghosts. The
problem with the usual approach of forgeting the higher mode spacetime
backgrounds and hoping that this is somehow equivalent to gauge fixing the two
dimensional w-gravity fields to zero, is that this is actually not possible.

Just to motivate the particle lagrangians introduced in the following section,
let me write down some non-trivial term in the W-gravity sigma model for the
bosonic string:
\begin{equation}
\int d^{2} \xi \sqrt{g} h^{\alpha \beta \gamma} (\partial _{\alpha} X^{\mu}
\partial_{\beta} X^{\nu} \partial _{\gamma} X^{\rho} G_{\mu \nu
\rho}(X(\xi))+\partial_{\alpha} \partial_{\beta} X^{\mu} \partial_{\gamma}
X^{\rho} \tilde{G}_{\mu \rho}).        \label{eq:wgrsigma}
\end{equation}

We need this kind of term in the sigma model, to accomodate string field theory
redefinitions according to

\begin{equation}
A=(G_{\mu \nu \rho}' a_{-1}^{\mu} a_{-1}^{\nu} a_{-1}^{\rho}+\tilde{G}_{\mu
\rho}' a_{-2}^{\mu} a_{-1}^{\rho}) |0 \rangle  .      \label{eq:wgrbrs}
\end{equation}

The BRST approach is consistent only if $Q_{A}^{2}=0$, which in the sigma model
language means that $A$ is a (1,1) conformal dimension primary field.

We will see that this kind of vertex implies that the moduli space of the
higher spin world sheet fields couples to the theory in such a way as to split
the string , and therefore changes the ghost number assignement of the second
quantised string field theory.

In the existing literature (up to my knowledge) there seems that there is the
feeling that $w_{\infty}$ is a kind of universal symmetry for all CFT, and that
it has something to do with discontinous deformations of the string,  and that
therefore a $w_{\infty}$-symmetric theory would somehow imply summation over
all genera (\cite{kn:ger}).  In more modern terms one could possibly say that
$w_{\infty}$ and its quantum extensions are the operator analogue of the
homotopy associative algebras defined on string fields.

Also this approach to string field theory will most likely solve the problem of
background independence. It is known that when we shift the background around
which we define our second quantised fields, we can define a new BRST operator
(as indicated above in the open case). One can then prove that this BRST
operator is nilpotent iff the background is on-shell, so described by a
conformal field theory. One would like off cource to know the lagrangian of the
theory around any background, not necessarilly conformal. A possible solution
to this problem could be the fact that we were naive to begin with to consider
the w-gravity sigma model with the higher spin w-gravity gauge fields gauge
fixed to zero. If we consider the `bigger' BRST coming from the w symmetry this
problem is likely to be resolved.

Another very important expectation in string theory that we will see becomes
now even more plausible is the conjecture (proven modulo very specific
assumptions) that there is a temperature (Hagedron temperature) that above
which the string is practically frozen in the sense that it has far less
degrees of freedom than one would naively expect the reason being that the
string exhibits a bigger symmetry. We claim in this paper that this mysterious
symmetry is $\dot{w}_{\infty}$. This symmetry resembes a lot the usual
$w_{\infty}$, but has a much clearer twistor interpretation.

\section{Contents}
Let us now summarize the contents of this work.

In chapter two and in appendix B the w gravity particle action was constructed
and gauge fixed in a non unitary gauge. As a result ghosts proliferate. The
higher order vertices of the theory can be constructed by using a geometric
identity inductively from lower point vertices. This results to Lorenz
covariantization of the action. Integration over the  $\dot{w}_{\infty}$
foliation ghosts should be equivalent to integrating over the Teichmuller space
of the graph, as explained in principle in chapter 4, but the explicit
calculation has not yet been performed. A conjecture about the existence of a
isomorphism between homotopy assosiative algebras and $\dot{w}_{\infty}$
algebras is made plausible, the connection between them being the geometric
identity.

In chapter three we repeat the same argumentation for the case of the bosonic
string. The w gravity gauge fields provide generalised Beltrami differentials
acting on  tensor products of the cotangent space of the Riemann surface. This
is exactly analogous to the situation encountered to the homotopy assosiative
algebras, where higher order products are not induced by lower order products.
The twistor space of the w gravity sigma models is constructed as being the set
of orbits of the fields of the theory under the $\dot{w}_{\infty}$ algebra,
with physical interpretation that they label the inequivalent second quantised
vacua. The existence of $\dot{w}_{\infty}$ symmetry in the sigma model action
imposes stringent constraints in the target space fields, which is an expected
phenomenon occuring at the topological phase of string theory. At the first
level these constraints allow the solution of the compactified components of
the fields as a function of the uncompactified ones. It is also argued that in
broken
 case when the leading cubic term is kept we can naturally compactify the
theory to four dimensions and also have a residual $SU(3) \otimes SU(2) \otimes
U(1)$ compactified group manifold.  In the topological phase pressumably we can
gauge away all the target space fields, and be left only with the foliation
ghosts, as a topological field theory. Finally a specific point in the twistor
space is exhibited where it exhibits tachyonic behaviour, as expected in the
inflation phase of the universe. This point is not written on compactified
time, therefore it is not Lorenz invariant, which is not nessesarily a drawback
because in the strongly coupled regime inertial frames might interact.

In chapter four we explain the role of foliations on the world sheet, and the
fact that summing over foliations is equivalent to summing over complex
structures, so therefore integrating over foliation ghosts is equivalent to
integrating over the Teichmuller space. It is also shown that the tachyon in
the open string case has its origin to coupling to closed string tadpoles. The
connection between the second and first quantised twistor spaces is exhibited.
{}From a general argument it is seen that that when we truncate string theory
to a single mode we expect the w algebra to trivialize.

In chapter five we explicitly show that indeed the w algebra trivializes, with
triviality established by gauge parameter redefinitions. Also in the spirit of
geometric identities we show that we can construct the four point Yang-Mills
vertex as the boundary of the collapse of two three point vertices along a
propagator collapsing along a non physical YM gauge particle. So therefore the
triviality of the w algebra is interpreted as generation of non physical modes
at the interaction points. We also explain the geometry of this construction,
which is in a sense a universal geometry over the group manifold, and we see
that it naturally exists in four dimensions.

Finally in chapter six we repeat the same techniques to the case of BRST open
string field theory. In the totally foliation gauge fixed theory of Witten we
expect and indeed find a trivial realisation of w symmetry, which allows us to
gauge away (say) half of the string. This theory is to be viewed as a gauge
fixed version of some formulation of string field theory which however includes
all possible foliations and therefore all possible ghost numbers for the second
quantised string fields, which is the expected topological phase. If we
partially gauge fix this theory (as we do in Witten's theory) we expect to have
to introduce foliation ghost action. A particular term of this action that can
help solve the background independence problem is exhibited. Also by using the
geometric identity we show that closed string field theory fields form a
representation space for $\dot{w}_{\infty}$.

\section{Particle in W-gravity background}
Let me begin by addressing the same issues in a simpler context, namely the
particle in a arbitrary W gravity background. Let me start with the simplest
case possible, a particle living in a one dimensional target space, where all
the W gravity currents are set equal to unity, and we keep only the W-gravity
gauge fields $h_{i}(\tau)$.
Consider the action
\begin{equation}
S=\int_{0}^{1} d \tau \sum_{i=2}^{\infty} h_{i}(\tau) \dot{x}(\tau)^{i} .
\label{eq:wpartact}
\end{equation}

Note that this action involves all the vertices to generate any
first-derivative interaction pattern. Soon it will be clear that due to the
gauge fixing of the gauge fields, not all these vertices have the same
statistical weight. Actually, by considering the coupling of this model to
external fields, it is strongly suggested that only the three point coupling
should be considered, so therefore reducing the symmetry to $w_{3}$. Remember
that this is also the case for covariant open string theory, and for light cone
closed string field theory,  but not of covariant closed string field theory,
suggesting that for at least the latter, truncation is not a good idea. Now,
after this small digression, let's return to our toy model.

This action has the obvious reparametrisation-like invariance
\[\delta h_{i}(\tau)=\frac{d}{d \tau}(h_{i}(\tau) \delta \tau),\]
\[\delta \dot{x} (\tau)= \delta \tau \ddot{x} (\tau),\]

\[[\delta_{1},\delta_{2}]=\delta_{\delta \tau_{2}
\stackrel{\leftrightarrow}{\frac{d}{d \tau}} \delta \tau_{1}},\]
under which the lagrangian is transforming by a total derivative as usual.

Note that the law of acting on the $\dot{x}$ is not the one induced by the
usual
\[\delta x=\delta \tau \dot{x},\]
but however that it differs by a term $-\dot{\delta \tau} \dot{x}$ which is
just a $\delta _{1}$ transformation (see (\ref{eq:partdiff})). Alternatively,
one could consider a non-local transformation law for the x, as induced by the
one on its derivative, and still preserve the algebra.

 In addition there is a $ w_{\infty}$-like symmetry, which we will denote by
$\dot{w}_{\infty}$, given by an infinite number of generators labelled by $k
\geq 1$ as follows (see fig. 1):

\begin{equation}
\delta_{(k)} \dot{x} =\delta \tau _{(k)} \frac{1}{c^{k-1}} \dot{x}^{k}  ,
\label{eq:wparttra1}
\end{equation}

\begin{equation}
\delta_{(k)} h_{n}(\tau)=- \theta (n-k-1) \: (n-k+1) h_{n-k+1} \delta
\tau_{(k)} ,  \label{eq:wparttra2}
\end{equation}

\begin{equation}
\delta _{diff} =\delta + \delta _{1} (\delta \tau _{1} =\dot{ \delta \tau}) .
\label{eq:partdiff}
\end{equation}

Note that this symmetry involves a dimensionful parameter with units of
velocity. It will turn out to be the velocity of light! From now on it will be
set equal to one.

These transformations with generators $J_{k}$ are easily seen to form an
algebra:
\begin{equation}
[J_{k},J_{l}]=(l-k)J_{k+l-1} . \label{eq:partalg}
\end{equation}

This algebra almost looks like the $t^{2}$ subalgebra of the usual
$w_{\infty}$, with commutation relations
$[t^{2}_{m},t^{2}_{n}]=(n-m)t^{2}_{n+m}$.

Also
\begin{equation}
[\delta ,\delta_{(k)}]=\delta_{(k)(\delta \tau_{k}'=-\delta \tau \: \dot{\delta
\tau_{k}})}, \label{eq:ideal}
\end{equation}

and
\begin{equation}
[\delta _{diff} ,\delta _{(k)}] =0 .    \label{eq:partcomm}
\end{equation}

Obviously the $\delta _{(k)}$ generators form an ideal, they act as raising
operators with respect to $\delta$ and they commute with $\delta _{diff}$,
which means that they do not change the mass of the string, in accordance to
the intuition provided also in \cite{kn:cap}, \cite{kn:wit}

Note that this symmetry is the immitation of the one boson field represantation
of the usual $w_{\infty}$, however it acts locally on the derivatives of the
fields and not the fields themselves therefore it is twistor inspired  and also
that it transforms only what would be called flag manifold coordinates, in
accordance with the standard geometrical interpretation \cite{kn:ger}.
\footnote{Also note the perplexing similarity of the $\theta (n-k-1)$ term in
equation~\ref{eq:wparttra2} with the corresponding connectivity degree of
Stiefel manifolds, which might not be a coincidence.}

The mental picture behind this symmetry is that a particle and all the random
lattice networks produced due to arbitrary splittings of the particle due to
interactions, are gauge equivalent entities (to the extent that the symmetry is
not broken, which of cource is the case due to the spacetime dependence of the
background, as will be seen shortly). Usually one puts in the definition
equation of the transformation law ~\ref{eq:wparttra1} a 1/k factor.
\begin{equation}
\bar{\delta _{k}}=\frac{1}{k} \delta _{k}
\end{equation}
 This is actually essential as we will see shortly, and it basically means that
the first quantised Hilbert states produced after the splitting of the particle
are indistinguishable, and weighted equally, and that in a two dimensional
plane (1+1) we have a symmetry factor of n (pure n point rotation ) for a n
point vertex, however most of the time I will omit it for notational
simplicity.

Let us now come to the crux of this section: By starting with the Newtonian
action (on the line) and generalising in a $\dot{w}_{\infty}$ inspired, modular
invariant way, we can reproduce the world-time gauge fixed massive Lorentz
action in two-dimensional spacetime. The idea basically is that an interaction
vertex is thought of as being composed from lower point interactions and
propagators. The analogous idea in string field theory is to perform a
Whitehead operation on the critical graph of a quadratic differential (see
chapters 4,5). In the $\dot{w}_{\infty}$ language this Whitehead operation is
thought of being the result of the action of a $\dot{w}_{\infty}$ current. Also
we will demand modular invariance, which we will allow us to weight equally the
graphs that arise after the Whitehead operation has been performed.

We start of cource with the action (for unit mass)

\begin{equation}
S_{Newton}=\int dt \frac{1}{2} \dot{x}^{2} .   \label{eq:newton}
\end{equation}

In the case of a toroidally Euclidean time  (\cite{kn:ati}) compactified
dimension, compactified time meaning perhaps that zero temperature and Hagedorn
temperature somehow are are therefore physically equivalent (and note that we
returned to the convensional uniform weight for the w generators, which will
actually it will turn out to be essential in what follows) this action is
invariant under the ideal of the odd generators :

\begin{equation}
\bar{\delta} _{2k+1} S_{N}^{tor. \: comp.} =0  .  \label{eq:npoddideal}
\end{equation}

Now let us consider the the moduli space of all graphs, in the same exactly
spirit with the geometric identity of closed string field theory \cite{kn:zwi}
(so therefore I will not explicitly repeat the arguments).  Although we could
work with stubs around the vertex, let us simplify the discussion by
considering a n-point vertex $V_{n}$ to be a point with a tensor product of
first quantised states assosiated with it. Then

\begin{equation}
 V_{n}=\partial V_{n} =\sum _{\begin{array}{ll} all \:  Whitehead \\ full
\:decompositions \end{array}} \partial _{p} R_{n-1}  ,  \label{eq:geoide}
\end{equation}

where $R_{n-1}$ is the set of all fully decomposed graphs with n external
lines, which therefore have n-1 internal propagators.

If a planar (planarity being used to make the combinatoric factors 1/n as
opposed to 1/n!) discrete graph with 2k+2 external lines and no internal
propagators is fully decomposed to a Whitehead equivalent graph with three
point vertices, it is easy to see that the number of vertices is 2k, and the
number of propagators is 2k-1.

\begin{tabbing}
\= external lines \= =2k+2  \\
\> propagators    \> =2k-1  \\
\> vertices       \> =2k
\end{tabbing}

Now let us examine $S_{2k+2}$ (or to be exact $h_{2n+2}$) (see fig. 2). This is
meant to be constructed out of a $S_{2k}$ and the $S_{2}$ graph by attaching a
common propagator to each in all possible ways and then shrink it\footnote{This
is probably equivalent to treating the one dimensional metric as quantum field,
with standard propagator, and calculating tree level corrections, or even
higher loop, to see that the one dimensional gauge fields condensate to the
values that appear in the Einstein action. See appendix B for more on this
point.} . We can choose one (out of 2k-1) internal propagators, and put the
extra line in two ways (either side of the propagator that lives in a plane
(1+1))  and take the uniform average for the  2k+2 external momenta. It follows
that
\begin{equation}
S_{2k+2}=\frac{2k-1}{2k+2} 2 S_{2k} S_{2}  .   \label{eq:iteration}
\end{equation}

By using the initial condition $S_{2}=1/2$ we get up to a additive costant
which we take to be -1, and a multiplicative constant that we choose to be -1
for positive energy

\begin{eqnarray}
S=\int dt (-1+\frac{1}{2} \dot{x}^{2}+\sum _{k=2}^{\infty} \frac{1 \times 1
\times 3 \times \ldots \times (2k-3)}{2 \times 4 \times 6 \times  \ldots \times
(2k)} \dot{x}^{2k}) = \nonumber \\
 = -\int \sqrt{1-\dot{x}^{2}} dt  , \label{eq:taylor}
\end{eqnarray}

which is simply equal to the taylor expansion of the Einstein action!

The above procedure we will christen w-completion, or w-contraction. To
summarise, it corresponds basically to condensation of world sheet gauge
fields. This gives a natural gauge fixed  action. This is a natural procedure
when we do not couple the particle to high spins. We also saw that when we try
to work in a general gauge, the lagrangian BRST gets very complicated. This is
in essence however the problem of background independence of string theory.
Hopefully the hamiltonian BRST will be more tractable. At the end of this work
we comment on the Hamiltonian framework.

Also note the following correspondence

\begin{equation}
Homotopy \: assosiativity  \longleftrightarrow  Geometric \: identity
\longleftrightarrow\dot{w}_{\infty} \: algebra           \label{eq:conjecture}
\end{equation}

This potential isomorphism between homotopy assosiative algebras and
$\dot{w}_{\infty}$ algebras will be made even more plausible in the next
sections.

\section{The bosonic string in w-gravity background}
The same ideas applied in the previous section to the particle will now be
applied to a restricted bosonic string model. What follows is a  generalisation
of \cite{kn:ber}, \cite{kn:hul1}, \cite{kn:hul2} in a way that will allow a
twistor interpretation and probably can be more easily treated in a Hamiltonian
formalism.
Consider the following action:

\begin{equation}
S=\int d^{2} \xi \sqrt{g}( g^{\alpha \beta} T_{\alpha \beta}+g^{\alpha \beta
\gamma}T_{\alpha \beta \gamma}+ g^{\alpha \beta \gamma \delta}T_{\alpha \beta
\gamma \delta}+infinite \; more)   ,  \label{eq:wstract}
\end{equation}

where
\begin{eqnarray}
T_{\alpha \beta}=\delta_{\mu \nu} \partial_{\alpha}X^{\mu}
\partial_{\beta}X^{\nu}, \nonumber \\
T_{\alpha \beta \gamma}=d_{\mu \nu \rho} \partial_{\alpha}X^{\mu}
\partial_{\beta}X^{\nu} \partial_{\gamma}X^{\rho} , \label{eq:defw}
\end{eqnarray}

and similarly for the higher order terms.

Obviously the symmetry properties of $d_{\mu_{1} \ldots \mu_{k}}$ must be taken
to be the same as $g^{\alpha_{1} \ldots \alpha_{k}}$ for the same k.

To motivate the above expression, this ( w-gravity non-linear sigma model)
action is supposed to describe a first quantised string propagating in
spacetime background fields $\delta_{\mu \nu}$,$d_{\mu \nu \ldots}$. The
topological obstructions in gauging away the two dimensional gauge fields
should correspond to interaction points of the string with some background,
with simultaneous splitting of the string, according to the order of
interaction as it appears in the action. So trerefore including w-gravity gauge
fields captures not only a particular Riemann surface, but also all possible
foliations on every Riemann surface (see section 5), so therefore the universal
Teichmuler space, in accordance with the conjecture of \cite{kn:ger}.

Under two dimensional reparametrisations the two dimensional fields transform
as their indices indicate. This is of cource a symmetry of the action. However
it is convenient to decompose the usual law for the derivative of the string
coordinate into two parts, a $\delta_{1}$ transformation (see below), and
consider the remaining part separately:
\[\delta(\xi^{\alpha})\partial_{\alpha}X^{\mu}=\xi^{\beta} \partial_{\beta}
\partial_{\alpha} X^{\mu}.\]
Then the algebra of these transformations closes as usually:
\[[\delta (\xi^{\alpha}),\delta (\eta^{\alpha})]=\delta (\zeta ^{\gamma}=\eta
^{\beta} \partial_{\beta} \xi^{\gamma} -\xi ^{\beta} \partial_{\beta}
\eta^{\gamma}).\]
One can also investigate if the above lagrangian is invariant under the
following transformation rules:

\begin{equation}
\delta_{k} (\partial_{\alpha} X^{\mu})=d^{\mu}_{\mu_{1} \cdots \mu_{k}}
\partial_{\alpha_{1}} X^{\mu_{1}} \ldots \partial_{\alpha_{k}} X^{\mu_{k}}
l_{\alpha}^{\alpha_{1} \cdots \alpha_{k}}(\xi)  ,       \label{eq:split}
\end{equation}

\begin{equation}
\delta_{k} g^{\alpha_{1} \cdots \alpha_{p}}=-(p-k+1) \theta (p-k-1)
g^{\alpha_{1} \cdots \alpha_{p-k} \alpha} l_{\alpha}^{\alpha_{p-k+1} \cdots
\alpha_{p}},            \label{eq:split2}
\end{equation}

where again we suppress powers of string tension.

For consistency we must demand that (\cite{kn:siepri})
\begin{equation}
\epsilon^{\beta \alpha} \partial_{\beta} \delta_{k} (\partial_{\alpha}
X^{\mu})=0 ,  \label{eq:consist}
\end{equation}

 which in the conformal gauge is easily seen to be satisfied provided that
\begin{equation}
l_{-}^{\; - \ldots -}=l_{-}^{\; - \ldots -}(\xi ^{-})  \; \; \; l_{+}^{\; +
\ldots +}=l_{+}^{\; + \ldots +}(\xi ^{+}) ,   \label{eq:solconsist}
\end{equation}

and the other l's are zero, and also if the theory has a good left-right
factorisation: $\partial _{+} \partial_{-} X=0 $, which in the case of closed
string equates the left and right Kac-Moody modes. The transformation laws
(\ref{eq:split}), (\ref{eq:split2}) have the natural interpretation of
splitting a first quantized string into k ones, provided that the left and
right modes do not mix. So therefore at the cases where it is possible to have
the above symmetry, the respective foliation pictures can be gauged away. To
make the above statement more clear, let us consider the following interaction
in complex notation
\begin{equation}
S_{int}=\int dz d \bar{z} \sum_{k \geq 3} \underbrace{ \partial X \ldots
\partial X}_{k \; terms} \frac{z^{k-1}}{\bar{z}}    ,
\label{eq:conformal}
\end{equation}

so therefore
\begin{equation}
g^{\overbrace{+ \ldots +}^{k \: times}}= \frac{z^{k-1}}{\bar{z}} ,
\label{eq:conformal2}
\end{equation}

and comparing with equation (\ref{eq:split2}) we see that
\begin{equation}
 l_{+}^{\; + \ldots +}=z^{k-1} ,  \label{eq:quadr}
\end{equation}

which is exactly the quadratic differential with a critical graph of k+1 order
(k+1 strings (say one incoming and k outgoing ) meet at a point).
This motivates the following geometric interpretation of the gauge fields (see
also \cite{kn:ger}): they are simply generalisations of the notion of Beltrami
differentials to the tensor product of the cotangent space of the Riemann
surface:
\[dz \longrightarrow dz+\mu^{z}_{\bar{z}} d \bar{z},\]
\begin{equation}
dz \otimes dz \longrightarrow dz \otimes dz + \mu^{z}_{\bar{z}} (dz \otimes d
\bar{z} +symm) + (\mu^{z}_{\bar{z}})^{2} d \bar{z} \otimes d \bar{z} +
\underbrace{\mu^{zz}_{\bar{z}} d \bar{z} +more}_{new \: terms}  .
\label{eq:homassos}
\end{equation}

\bigskip

We see that in the tensor product of two spaces the product is not induced from
the product of the space itself. This is highly reminiscent to the situation of
homotopy assosiative algebras of string field theory.

The above models are tractable analytically. At this point note that the
equations of motion we obtain for them are non-linear, so the high order
oscillator modes are frosen in the sense that we dont have arbitrary linear
coefficients (to be compared to equation (\ref{eq:spaceconsist})). At the end
of this section we will present a specific solution.

Let me commend a bit more on equation (\ref{eq:split}). One can define two dual
spaces: The space $T$  of tensor fields on the world sheet  and the space of
tensor fields on target space $T'$, the duality being provided with the natural
pullback as follows:

\begin{equation}
d^{(\ast) \, \mu}_{\alpha_{1} \ldots \alpha_{k}} \equiv d^{\mu}_{\mu_{1} \ldots
\mu_{k}} \partial_{\alpha_{1}} X^{\mu_{1}} \ldots \partial_{\alpha_{k}}
X^{\mu_{k}}  ,   \label{eq:dual}
\end{equation}

the duality being natural if and only if the equation of motion is obeyed for
the X's (and also remember that only then the transformation law is
consistent). With this notation the equation (\ref{eq:split}) is simply written
as

\begin{equation}
\delta_{k}(\partial_{\alpha} X^{\mu})=d^{(\ast) \, \mu}_{\alpha_{1} \ldots
\alpha_{k}} l_{\alpha}^{\alpha_{1} \ldots \alpha_{k}} .
\label{eq:twisplit}
\end{equation}

Equation (\ref{eq:twisplit}) is highly reminiscent of a twistor like-split for
the velocities (actually the appearence of momenta would be probably more
desirable), see \cite{kn:pen} . The appearance of this kind of structure was
observed (in a different guise) in \cite{kn:witb}, \cite{kn:twi}. Shortly I
will try to show that the mysterious (\cite{kn:gro}, \cite{kn:eva}) symmetry of
the string is manifest in the twistor space of the string (to get a clear proof
we must actually go to the phase space of the string, and this will hopefully
be the subject of a future work).

One can calculate (provided that the d fields are constants) that
\begin{equation}
[\delta (\xi),\delta _{k} (l_{\alpha}^{\alpha_{1} \cdots
\alpha_{k}})]=\delta_{k}(l'=-\xi^{\beta} \partial_{\beta} l),
\label{eq:stringideal}
\end{equation}

so still the $\delta_{k}$ operators form still an ideal in the same way as in
the particle case ~\ref{eq:ideal}, and therefore commutes with the target space
diffeomorphisms (eq.(\ref{eq:partcomm}))

The calculation of the commutator of two w-generators is more involved. If it
so happens that
\begin{equation}
d_{\mu_{1} \ldots \mu_{k} \mu_{k+1}}d^{\mu_{k+1}}_{\mu_{k+2} \ldots
\mu_{k+l}}=d_{(\mu_{1} \ldots \mu_{k+l})} , \label{eq:spaceconsist}
\end{equation}

(where the parenthesis denotes total symmetrisation), for example if $d_{\mu
\nu \ldots}$ is equal to one if and only if all the indices are the same, the
the algebra closes in the simple and by now familiar way. Unfortunately
(perhaps) it is not satisfied for the symmetric symbol of SU(3).

In the case that the above condition holds, (and note that this is a condition
in the $T'$ space) we will say that the string lives in the $\dot{w}_{\infty}$
radius. In this case there is a huge gauge symmetry that trivialises the higher
modes of the string, in accordance with the intuition provided in
\cite{kn:gro}, \cite{kn:eva}.
However, there is anoter case of interest, namely when only $\delta_{1}$ and
$\delta_{2}$ are independent (and note that they form a subalgebra too!), and
for the rest the following condition holds (and note that this is a condition
in {\em T} space):

\begin{equation}
l_{\alpha}^{\alpha_{1} \ldots \alpha_{l-1}
\alpha_{l}}l_{\alpha_{l}}^{\alpha_{l+1} \ldots
\alpha_{l+k}}=l_{\alpha}^{(\alpha_{1} \ldots \alpha_{l+k})}.
\end{equation}

In this case we will say that the string lives in the $w_{3}$ radius (note that
equation (\ref{eq:quadr}) satisfies this relation). Note also that this case is
more natural for open strings, because if we believe Witten's open string field
theory three string interactions is all it matters (when the background is
trivial), which is generated by $\delta_{2}$.

We can consider three scenarios for explaining the reduction of the theory to
four physical dimensions, the last one being our favorite.

 In the first scenario we  consider the fundamental representation of SU(3)
(six real dimensional) to provide the compactified dimensions of the
superstring, as follows:
\[X_{L}=\underline{3} \oplus \underline{8} +X_{observed},\] \nopagebreak
\[X_{R}=\underline{3}^{\ast} +X_{observed}.\]

The above equations mean that  one can gauge away all dimensions except four,
in the above simple model. The way that this gauging away is to be accomplished
is no clear in the Lagrangian formalism we consider, however note that
$l_{\alpha}^{(\beta \gamma)}$ has six components and $l_{\alpha}^{(\beta \gamma
\delta)}$ has eight components. Now a consistency test (which can be thought of
as a second scenario) is  that we do not alter first quantised states by the
action of the symmetry (remember that $\dot{w}_{\infty}$ symmetry is basically
an artifact, providing gauge equivalent portraits of the a single particle
being split, however the split portrait can basically be gauged away. In the YM
case we could say that a physical gauge particle can emit a longitudinal gauge
particle, and this is an iteraction, which however is trivial).
So if D is the dimension of target space for the totally foliation gauge fixed
theory, and $ D'$  in the $w_{3}$ topological phase, then (by fixing $\tau$
reparametrisations using the target space time)

\begin{eqnarray}
D-1=(D'-1)^{2},  \\  \label{eq:xy2}
D=1 \leftrightarrow D'=1 , \nonumber \\
D=10 \leftrightarrow D'=4,  \\  \label{eq:xy10}
D=26 \leftrightarrow D'=6 . \nonumber
\end{eqnarray}

 So if we want the foliation split to be just a gauge artifact, then we
actually need $D'=4$. Note also that D=1 is a fixed point.

Now let us come to the third and most well understood scenario. The reader at
this point must read appendix \ref{ap:cov}. We treat the matter content of the
sigma model as elements of a group manifold. We know that at the quantum level
two open strings can be combined to form a closed string, or in a theory
containing both open and closed strings this vertex (COO) exists even at the
classical level . We demand that (see fig.3)
the closed string (propagating in the trivial background) lives in the adjoint
representation of some group manifold with dimension 24, for example SU(5), and
we also have a phase playing the role of a Teichmuller parameter for the
spliting of a closed string of length $2 \pi$ to open strings of length $\pi$,
and a dilaton field with central charge one to kill the conformal anomaly.  In
the two open string sector we demand that each open string lives in the
fundamental representation of the same group, and we also need one more dilaton
of the same central charge (one) for the one more string we get after the
splitting. SU(5) still works as $5^{2}+1=26$. In the light cone gauge we would
simply say that the transverse degrees of freedom of the closed string form the
adjoint representation of SU(5) and also we have a phase as a modular parameter
for the splitting to two open strings. The open string sector lives in the
fundamental representation of SU(5):$ 24+1=5^{ 2}$. This consideration leads us
to believe th
at SU(5) must be of some relevance.

Now let us work in the extended models in arbitrary background. The 27+1
dimensional manifold we consider to provide the totally symmetric cubic
coupling in (2):
\begin{equation}
G_{\mu \nu \rho}=d_{\mu \nu \rho},
 \end{equation}
is

\begin{equation}
\overbrace{SU(5) \otimes R^{3}}^{27} \otimes \overbrace{\phi}^{-1} ,
\label{eq:4d}
\end{equation}

where the grand unified group SU(5) describes the transverse coordinates, and
also

\begin{equation}
SU(5) \rightarrow \overbrace{SU(3) \otimes SU(2) }^{8 \times 3} .
\label{eq:4d2}
\end{equation}

The three commuting generators of $R^{3}$ along with the dilaton mode provide
our four dimensional world. The transverse coordinates of the string are
coupled in the sigma model in a cubic way with the invariant symmetric tensor
of SU(5), therefore indicating that at the field theory level this should
correspond to a self-interacting gauge theory based on the same group. The
non-self-interacting U(1) gauge field is introduced in the sigma model as usual
in a linear way:

\begin{equation}
\int d^{2} \xi A_{\mu} \partial _{+} X^{\mu} . \label{eq:lin}
\end{equation}

Let us now look at the $\dot{w}_{\infty}$ radius equation
(\ref{eq:spaceconsist}) at the first three levels combined with the above
compactification.

\begin{equation}
A^{\hat{\mu}} G_{\hat{\mu} \hat{\nu} \hat{\rho}}=G_{\hat{\nu} \hat{\rho}}.
\label{eq:firslev}
\end{equation}

We apologize for switching temporarily notation to conform with the recent
literature. The indices take the following values:
\[\hat{\mu}=0,\ldots ,D+d-1  \;\;\;\;\; \mu =0, \ldots ,D-1 \;\;\;\;\; \alpha
=1, \ldots ,d ,\]
with d compactified dimensions.

We use  the ansatz

\begin{eqnarray}
A_{\hat{\mu}}=(A_{ \mu},A_{\alpha}), \\
G_{\hat{\mu} \hat{\nu}}=\left( \begin{array}{ll}
g_{\mu \nu}+A_{\mu}^{\alpha}A_{\alpha \nu}+A_{\mu \nu} & A_{\mu \beta}\\
A_{\alpha \nu} & g_{\alpha \beta}
\end{array}
\right) ,\\
G_{\mu \nu \rho}=0 \;\;\;\;\; G_{\alpha \beta \gamma}=d_{\alpha \beta
\gamma},\\
G_{\mu \nu \alpha}=a g_{\mu \nu}A_{\alpha}+b A_{\mu \alpha} A_{\nu}, \\
G_{\mu \alpha \beta}= c g_{\alpha \beta} A_{\mu} + d A_{\mu \alpha} A_{\beta}.
\end{eqnarray}

By substituting the above equations in ( \ref{eq:firslev}) we can solve for the
components of the fields along the compactified dimensions and we also we
obtain a D-dependent constraint on the moduli space of the ansatz, as one can
easily prove.

 Now returning back to our original notation and to  eq.( \ref{eq:dual}), we
are lead  to believe that we must actually treat {\em T,T'} on equal footing,
and that actually the `Grand String Symmetry' is manifest on $T \times T'$.
This symmetry is broken by the condensation of the world sheet gauge fields. At
the regime where the 2d gauge fields are fluctuating, we have the Topological
Phase of string theory. In that phase the string is in a pure gauge state: By
equation~\ref{eq:split2} we can gauge away all the action, so the topological
phase is a zero action instanton phase. At the topological phase
$\dot{w}_{\infty}$ acts linearly on $T \times T'$, as it should according the
twistor ideas (remember that the twistor space for the usual particle is a
linear represantation space of the SU(2,2), whereas on the usual space SU(2,2)
is represented non-linearly. In the spirit of \cite{kn:twi} we define the
Twistor Space $TS$ of the theory as

\begin{equation}
TS_{sec.quant} \equiv \frac{T \times T'}{\dot{w}_{\infty}} =inequivalent \;
vacua .         \label{eq:twispace}
\end{equation}

In the above equation it is clear that the orbits of $w_{\infty}$ are the
generalisation of the notion of $\alpha$-plane. The above definition of twistor
space is second quantised one. Compare with the definition of the first
quantised twistor space (\cite{kn:twi}):
\begin{equation}
TS_{first.quant} \equiv \frac{DiffS^{1}}{S^{1}} =inequivalent \; complex \;
structures   .        \label{eq:twispace1}
\end{equation}

Probably the moduli space $TS_{first.quant}/w_{\infty}$ is relevant too,
perhaps as providing a mechanism for the discreet symmetries of the string,
because it basically counts singularities in phase space.

Now let us see what happens before the 2(1) dimensional gauge fields condensate
(in which case the vacuum expectation value of which defines a prefered gauge
fixing). Let us consider an arbitrary gauge fixing.
Let me be explicit and consider the following simplified action (in one
dimensional pure space target space, with + coordinate being the world sheet
time).

\begin{equation}
 S_{no-condenced} =\int d \xi ^{+} d \xi ^{-} \frac{\partial_{+} X \partial_{-}
X}{1+h^{+} \partial_{+} X} , \label{eq:winfstring}
 \end{equation}

where $h^{+}$ is taken to be a constant for simplicity. The equation of motion
factorizes simply (as in the particle case too):

\[(1+h^{+} \partial_{+} X)[\partial_{+} \partial_{-} X(1+h^{+} \partial_{+}
X)-2h^{+} \partial_{+}^{2} X \partial_{-} X] =0 .\]

The solution

\[\partial_{+}X =- \frac{1}{h^{+}}.\]

we can regard as providing initial conditions for the string, which in the
small $h^{+}$ limit is tachyonic. Note that

\begin{equation}
h^{+} \longrightarrow 0  \Longrightarrow \left\{ \begin{array}{l}
 S_{w_{\infty}} \longrightarrow S_{bosonic \; string}\\
\partial_{+} X \longrightarrow indefinite.
\end{array}
\right.
\end{equation}

Let me commend on the above limit. By totally gauge fixing the chiral
$\dot{w}_{\infty}$ string lagrangian we obtain the usual light cone bosonic
string lagrangian. However by forgetting how we obtained the bosonic string
lagrangian, we also forget not only the first order part of the equation of
motion, but also there might exist problems with the ghosts. These
$\dot{w}_{\infty}$ ghosts will be refered to as foliation ghosts. The fact that
we cannot simply forget about them, might be the key to the solution of the
background independence of string field theory.
Let us now investigate a model in the spirit of equation (\ref{eq:conformal}):
\begin{equation}
S=\int dz d \bar{z}( \partial X \bar{\partial} X + \frac{2}{3}
\frac{z^{2}}{\bar{z}} (\partial X)^{3}),    \label{eq:solvtoy}
\end{equation}

with equation of motion
\begin{equation}
\partial (\bar{\partial} X+\frac{z^{2}}{\bar{z}} (\partial X)^{2})=0  .
\label{eq:soltoy}
\end{equation}

The above equation has antiholomorphic solutions, and a specific combination of
$\frac{ln^{2} z}{ln b \bar{z}}$ and $\frac{lnz}{lnb \bar{z}}$. We see that the
higher modes of the string are frozen, as we also saw for the background
expectation values in equation (\ref{eq:spaceconsist}).
Let me for completeness write down the equation of motion (and of cource there
are also constraints ) too:
\begin{equation}
\partial_{\alpha}\partial_{\beta}X^{\nu}(\delta_{\nu}^{\mu} g^{\alpha
\beta}+3d^{\mu}_{\nu \rho} g^{\alpha \beta \gamma}
\partial_{\gamma}X^{\rho}+4d^{\mu}_{\nu \rho \sigma}g^{\alpha \beta \gamma
\delta} \partial_{\gamma} X^{\rho} \partial_{\delta}X^{\sigma}+more)
\label{eq:nest}
\end{equation}

We see appearing a nested structure of derivatives, phenomenon that we will
encounter in the string field theory too (\ref{eq:64}).

\section{Foliations on the world-sheet}

Consider a Riemann surface $\Sigma$ embedded in D-dimensional spacetime. It is
natural to consider the space of all foliations of this surface.The role of
foliations is twofold: Some foliations can be thought of as being induced by a
morse function T (global time) defined on the world sheet which is interpreted
as target space time restricted on $\Sigma$ (in particular $T=X^{0}$ is a
reasonable choice), so every leaf of the foliation is just the inverse image
under T (with T being a trivial transverse measure on the foliation) of some
target space time. Then the critical points of the foliation coincide with the
critical (zero Hessian) points of T. The sum of the indices assosiated with
these critical points is just the Euler number of $\Sigma$ which is consistent
with the interpretation that they correspond to interaction points in the
interacting theory (points that all the curvature of $\Sigma$ is concentrated,
giving a curvature singularity). So the usual problem of the geometric
interpretation of intera
ctions in point particle theories has appeared again in a different guise. The
fact that by doing a conformal transformation on the (Wick rotated) $\Sigma$ we
can move the curvature singularity around, or even smooth it out, is translated
in the other language by saying that we can move the foliation singularities
around by doing diffeomorphisms, and also we can do Whitehead operations
(fig.4) on $\Sigma$ to split the singular points of the foliation (keeping of
course the sum of the corresponding indices the same) which is equivalent to
redistributing the curvature on $\Sigma$. This Whitehead operation clarifies
the fact that in light cone sting field theory only at most cubic terms of
interactions need be considered, as I will explain shortly.

Preverse things can happen if the normal vector field assosiated with T is
spacelike in the target space $\cal T$ sense. In this case the foliation of a
naively thought of open string is the one one would expect to see in a closed
string, as shown in figure 5 . This means that $\cal T$-space tachyonic
theories interpret open strings as being  closed string-tadpole theories. This
kind of foliation is not forbidden a priori due to the existence of the tachyon
in the spectrum of the (un-GSO projected as being anyway bosonic) string
theory, the conformal structure on $\Sigma$ being on-shell the same as the one
inherited from $\cal T$.
Not only that,but actually global diffeomorphisms ``vertical'' to $\cal F$
actually generate such foliations, thus explaining the geometrical origin of
the tachyon. In NSR case diffeomorphisms must preserve the spin structure (or
else we have overcounting because we eventually sum over spin structures), so
it is self consistent to consider only timelike T. This explains the geometric
origin of GSO projection.

This $\sigma \leftrightarrow \tau$ duality is of cource exactly the same
duality between P and X', in the canonical transformation sense, as is well
known.

Other strange case would be the one where some or all leaves are $\Sigma$ space
filling curves. Then, modulo a fundamental resolution lenght and a $\sigma$
reparametrisation this situation corresponds to closed strings intertwined
together in a complicated way. However this foliation cannot be induced by
global time T. This means that global gauge fixing of the world sheet time is
severely restrictive for the admissible underlying foliations, namely the set
of the global time foliations is a set of zero measure (in the appropriate
sense) in the set of all foliations. It is clear that this phenomenon has a
trivial counterpart even in ordinary gravity theory, where the non-existence of
global time is related to the fact that its existence would imply the existence
of a nowhere vanishing vector field (orthogonal to constant time curves) which
is in some spacetimes forbidden (depending upon the vanishing or not of the
Euler number).

 The other role of foliations is to provide a natural setting for deforming
conformal structures. Let me review the (well known in mathematics literature,
e.g. \cite{kn:ast}) argument. Fix a conformal structure on the surface.
Consider an arbitrary pair of transversal foliations, and the natural
coordinate system z=x+iy assosiated with them. Then w=x+iay defines a new
conformal structure on the surface. This obviously establishes the known fact
that the dimensionality of moduli space is two times the number of propagators
on the surface, equal to 6g-6+2n. This 2n is just an additional lenght and
angle for each external leg. The Whitehead operation is natural in the sense
that Whitehead equivalent foliations define the same deformation in the moduli
space.
This means that integrating over the set of conformally inequivalent two
dimensional metrics is the same as integrating over the Whitehead inequivalent
foliations, and we can gauge fix this symmetry in the action using Fadeev-Popov
foliation ghosts.

\begin{equation}
\int D[g] DX e^{S(X,g)} = \sum_{\begin{array}{ll} Whitehead \: non \:
equivalent \\ foliations \end{array}} \int D c_{fol} DX e^{S(X,c_{fol})} .
\label{eq:modint}
\end{equation}

The sum over the inequivalent foliations can be in principle performed using
the techique we exhibited in the particle case. The result there was just
Lorenz covariantisation. In the YM case we will see shortly that it is Gauge
covariantisation. In the string case it will be the seeked for `Grand String
Symmetry'.

 Also an important fact is that foliations with space filling curves of the
type mensioned before are needed in order to define points in the
compactification divisor of moduli space.
 After the foliation is performed one can still consider diffeomorphisms that
preserve $\cal F$ in the sense that they leave its leaves invariant. The
typical situation is as in figure 6: In figure 6 it is clear that indeed the
sum of the indices of the vector fields assosiated with diffeomorphisms keeping
the foliation invariant is indeed equal to $\chi$($\Sigma$) (in the orientable
case), both interaction points being hyperbolic,and the Euler number being -2.
Also it is clear that these diffeomorphisms belong to the group Diff($S^{1}$)
away from the interaction points. The leave passing through the interaction
point looks like F8(figure eight). The action on F8 is product of two
diffeomorphisms of $S^{1}$ which both leave the intersection point
invariant.Such diffeomorphisms have nessesarilly rational rotation number.

 So far I considered the configuration loop space. It is of interest to
consider also the phase space induced as a cotangent bundle on the loop space.
This space can be coordinised with the normal modes of the string and the
assosiated momenta. In the case of the open string I can perform the usual
redefinition $ x=p+dq/d\sigma$ to obtain periodic coordinates on the phase-loop
space PL, so PL is just the set of embeddings of $S^{1}$ in our spacetime. Lets
confine ourselves to based loops so that we ignore all the zero modes. On their
space bPL, one can define in a natural way a complex structure, with the
positive frequency modes being considered holomorphic coordinates. So as the
string vibrates, it traces out a curve in bPL, different curves in bPL
corresponding to different initial (first quantized) conditions.
On $S^{1}$ we have the natural action of $Diff(S^{1})$. What is magical about
this group is that its action is equivalent (up to homeomorphism conjugacy
class) to a pure rotation, characterized by some rotation number (which doesn't
affect the zero mode anyway). So up to conjugacy class the complex structure of
bPL is uniquely defined, and anyway even after a general diffeomorphism the
phase portraits of bPL will be topologically equivalent. This is a big
simplification that does not hold in finite dimensional case, e.g not every
SO(2d) transformation belongs to a SU(d) conjugacy class of a U(1)
transformation. Non-interacting open string theory is unique in this sense. The
physical interpretation is that (up to topological equivalence) the string can
tell which particle mode is which! (trivial physics corresponding to not
totally trivial mathematics).

Now let us investigate the correspondence between first and second quantised
twistor spaces, equations (\ref{eq:twispace}), (\ref{eq:twispace1}). We see
that $TS_{first.\: quant}$ is the set of symplectic structures on the phase
space (because a complex structure induces a symplectic structure), and every
symplectic structure corresponds to a topological conjugacy class of
diffeomorphisms. We also have seen that $\dot{w}_{\infty}$ corresponds to
changing the foliation picture of the first quantized theory, therefore to a
deformation of the complex structure, therefore a fiber over $\dot{w}_{\infty}$
corresponds to no deformation of the complex structure, therefore to a pure
rotation of the oscillator modes. So

\[ Second \: \: quantised \: \:  \dot{w}_{\infty} \: \: transformation \]
\centerline{$\updownarrow$}
\[ first \: \: quantised \: \: pure \: \: rotation . \]

Another way to say this, is by looking at the effect of a finite conformal
transformation on fourier modes over a loop:

\[\delta a_{m} =e^{p \cdot L} a_{m} e^{- p \cdot L} , \;\;\;\;\;\;\; p \cdot L
\equiv \sum p_{n} L_{n},\]

which is a  (infinite) polynomial over oscillator modes. Also by substituting
in (\ref{eq:split}) the oscillator expansion of the fields that appear there,
we see that the effect of a w generator on a oscillator mode $a_{m}$ is to
change it in a infinite polynomial way (homogeneous in k, to be more exact), as
indeed desired.

An important check of this would be to see what happens at the second quantised
level if we truncate the theory to just one level. If we do that then in the
one-complex dimensional space of the truncated string mode, only a pure
rotation exists (the plane has a unique complex structure). Therefore in the
second quantised level $\dot{w}_{\infty}$ must act trivially too! This we will
verify in the following section.

\section{$\dot{w}_{\infty}$ and Yang-Mills}
Since it is possible to see $\dot{w}_{\infty}$ at the level of particle/string
sigma model  actions, with the natural interpretation of a symmetry that gauges
away interactions, it is of high interest to see what this symmetry exists at
the level of second quantised particle-based theories, with the hope of
understanding better the role of gauge symmetry at the interacting (so
therefore quadratic and higher order in the gauge fields) level, We will see
that $\dot{w}_{\infty}$ is realised in a basically trivial way. This is
actually to be expected. We already saw that $\dot{w}_{\infty}$ is basically a
symmetry of the topological phase of string theory, where we can basically
gauge away all the oscilator modes of the string, and actually spacetime
itself. The reason that this was possible is the fact that there is an
isomorphism between the space of the world sheet gauge fields and the space of
spacetime fields, isomorphism which is natural on-shell. So basically we can
use spacetime fields to built gauge para
meters to gauge away spacetime fields. Now when we truncate the theory only to
one component field, say YM, or gravity (which is itself of YM type) we expect
the representation of $\dot{w}_{\infty}$ to this one field to trivialize it is
like forcing a group to be represented in a space of smaller dimension than its
defining representation. This is what actually happens: We can actually prove
that the natural candidate for such transformation laws indeed satisfies the
desired commutation relations, and it is indeed trivial up to field
redefinitions, yet tensor operator product inspired, as desired. Note that this
is not trivial, in the sence it is unique: There are no other redefinitions of
the gauge parameters that would close to form an algebra, let alone the desired
algebra. Note also the perplexing possibility of representing a group in a
descrite space of smaller dimension that the defining representation, which
might shed some light to the discrete symmetries of the string. One can also
see that the sam
e construction $(\delta_{k} \dot{x} =\frac{d}{d \tau} (\dot{x}^{k} \delta
\tau_{k}))$ does not work in the particle (or string) case, which is consistent
with the interpretation that $\dot{w}_{\infty}$ simply splits the particle
(string) without changing the legs qualititavely.

We will see that we can realise this symmetry in a way that seems naively
trivial due to field redefinitions, however at the quantum level the ghost
sector is different, and also new possibilities for spontaneous symmetry
breaking due to vacuum expectation values of Wilson loops arise. The relevance
of this mechanism to string theory was first noted in a different context in
\cite{kn:nan}.

Let me study the abelian case first. Consider the following transformation
laws:

\[\delta_{0} A_{\mu}= \partial _{\mu} L,\]
\[\delta_{1} A_{\mu}= \partial _{\mu} (A^{\nu} L_{\nu}),\]

\begin{equation}
\delta_{k} A_{\mu}= \partial _{\mu} (A^{\nu_{1}} \ldots A^{\nu_{k}} L_{\nu_{1}
\ldots \nu_{k}}).  \label{eq:winfem}
\end{equation}

Then it is easy to prove that the following algebra closes

\[ [\delta_{0},\delta_{1}]=\delta_{0}(L'=(\partial^{\nu}L)L_{\nu}),\]
\[ [\delta_{0},\delta_{2}]=2\delta_{1}(L'_{\nu}=(\partial^{\rho}L)L_{\nu
\rho}),\]
\[ [\delta_{1},\delta_{2}]=\delta_{2}(L'_{\rho
\sigma}=2(\partial^{\nu}L_{\sigma})L_{\nu \rho}-(\partial^{\nu}L_{\rho
\sigma})L_{\nu}))+\delta_{1}(L'_{\rho}=4\partial^{\nu}A^{\sigma}L_{[\sigma}L_{\nu ] \rho}),\]

If one chooses $L_{\nu \rho}=L_{\nu}L_{\rho}$ then in the above expresion
$L'_{\rho}=0$ and $L'_{\sigma \rho}=2L_{\nu}(\partial^{\nu}L_{[\sigma})L_{\rho
]}$ and the algebra closes as simply as the particle case, although it is non
really imperative that we demand this. However if we do,  we can give the
following intuitive interpretation for the role of this weird symmetry:
$\dot{w}_{\infty}$ is the generalisation in the (interacting) Fock space of
what gauge symmetry is in the (one particle) Hilbert space. The same phenomenon
will be observed in the non-abelian case (see below) and in the string field
theory (next section), however the detais are not fully clear to us. We
moreover see that actually in this case the algebra can be truncated:
[\delta_{1},\delta_{2}]=0$. This is good for the abelian case since if it was
not the case we would get a four point interaction term in the classical
action, as we will get it for YM (see below), and then we  would probably have
to add a multitude of w-ghosts.

Similarly one can  check that the algebra closes for the rest of the
generators:
\begin{eqnarray}
[\delta _{k} (\Lambda _{(k)}),\delta _{l} (\Lambda _{(l)})]=(l-k) \delta
_{k+l-1}(\Lambda _{(k+l-1)}=(\partial ^{\nu} \Lambda _{(l)})\Lambda
_{(k-1)\nu}-(k \leftrightarrow l) + \nonumber \\
+kl \delta _{k+l-2} (\Lambda _{(k+l-2)}= F_{\rho \nu} \Lambda ^{\nu}_{(k-1)}
\Lambda ^{\rho} _{(l-1)} A_{(k+l-2)}),
\end{eqnarray}
where $A_{(k)} \equiv A_{\mu _{1}} \ldots A_{\mu _{k}}$ and $\Lambda _{(k)}
\equiv \Lambda _{\mu _{1} \ldots \mu _{k}}$.
In the above equation we see that again that when the high order gauge
parameters are reduced to the (symmetric) product of lower order parameters
(order one ), or if the field is purely topological (zero gauge strength) then
the algebra is the same with the particle counterpart, otherwise there is one
lower order contribution (which vanishes for topological configurations).
In the non abelian case things work a little bit differently (in a way which is
strikingly similar to tensor operator algebras of \cite{kn:pop},\cite{kn:wit},
again in the same spirit of our considerations):

\[\delta_{1}A_{\mu}=[D_{\mu},L],\]

\begin{equation}
\delta_{k}A_{\mu}=[D_{\mu},[A_{\nu_{1}},[\ldots,[A_{\nu_{k-1}},L^{\nu_{1}
\ldots \nu_{k-1}}] \ldots ]]] .  \label{eq:winfym}
\end{equation}

Then one can easily compute (using the Jacobi identity) that

\[[\delta_{1},\delta_{2}]=\delta_{1}(L'=[\partial_{\nu}L,L^{\nu}])+\delta_{2}(L'^{\nu}=[L,L^{\nu}]) .\]

If $L_{\nu}=\partial_{\nu}L$ then only the second term survives to get the
particle-inspired result.

Now let's consider the next candidate transformation law:

\[\delta_{3}A_{\mu}=[D_{\mu},[A_{\nu},[A_{\rho},L^{\nu \rho}]]] .\]

Then

\[[\delta_{1},\delta_{3}]=\delta_{3}(L'^{\nu \rho}=[L,L^{\nu
\rho}]+\delta_{2}(L'^{\rho}=[\partial_{\nu}L,L^{\rho
\nu}])+\delta_{1}(L'=[\partial_{\nu}L,[A_{\rho},2L^{[\nu \rho]}]]) .\]

{}From the above equation we deduce that $L^{\nu \rho}$ is symmetric, as
expected since identical gauge particles must combine wavefunctions in a
symmetric state, and proportional to the unit matrix in the Yang-Mills indices
if we demand the algebra to close as in the particle case:
 \begin{equation}
L^{\nu \rho}=G^{\nu \rho}1_{YM}.   \label{eq:ymstr}
\end{equation}

Similar relationships hold for the higher modes too. Therefore the algebra
basically gets truncated down to the third level.

To show that this is desirable  let's see what it means to have $\delta_{2}$
non trivial. It means that we can split the gauge particle into two, however we
must keep the degrees of freedom unaltered, so one of them can be physical and
the other unphysical, or both of them can be partly physical. If we take the
three point interaction term in the YM action for granded and apply a  $\delta
_{2} $ transformation on an (external physical) gauge field, and interpret this
action as leading to a physical external field and a longitudinal one, and we
repeat the process two times, then we get the four point function with the
correct normalisation, as we did for the Lorenz particle! The details go as
follows (see fig. 7):

\[ L_{YM}=L_{2}+L_{3}+L_{4}, \: \: \: \: \; \;  L_{3}=-ig \partial _{\mu}
A_{\nu}^{a} f_{a}^{\: bc} A_{\mu b} A_{\nu c} , \]

\begin{eqnarray}
S_{3} \star S_{3} \equiv - \frac{1}{4} g^{2} \int \int dx \: dy f_{a}^{\: bc}
f_{d}^{\: ef} A_{\mu b}(x)A_{\nu c}(x) A^{\rho}_{e}(y)A^{\sigma}_{f} \times
\nonumber \\
<\partial _{\mu} A^{\nu a}(x) \partial_{\rho} A^{\sigma d}(y)>_{L} ,
\label{eq:ymcontr}
\end{eqnarray}

where in the above equation we only contracted the derivative (longitudinal)
terms (to get a modulary invariant result), and we inserted the factor 1/4 by
planarity in two dimensions (actually we can pretty much set it to whatever
value we want to by choosing appropriately the value of gauge fixing parameter
$\alpha$, which is not too bad because the gauge fixed theory is not gauge
invariant anyway), and the symbol $< \ldots >_{L}$ is defined as momentum
integration constrainted only on longitudinal momenta ($\frac{k^{\mu}
k^{\nu}}{k^{2}}=\delta^{\mu \nu}$):

\begin{equation}
<\ldots>_{L} \equiv \int_{\Pi _{\mu \nu}^{L}=\delta _{\mu \nu}} d^{D} k
(\ldots) , \label{eq:ymlong}
\end{equation}

so therefore
\begin{eqnarray}
<\partial _{\mu} A^{\nu a}(x) \partial_{\rho} A^{\sigma d}(y)>_{L} = \nonumber
\\ =\int_{L} \frac{d^{D}k}{(2 \pi)^{D}}(\frac{-i}{k^{2}}(\delta ^{\nu
\sigma}+(\alpha -1)\frac{k^{\nu} k^{ \sigma}}{k^{2}})\delta ^{ad}
\partial_{\mu} \partial_{\rho}' e^{-ik(x-y)}) =  \nonumber \\
=-i\alpha \delta_{\mu \rho} \delta^{\nu \sigma} \delta ^{ad} \delta ^{D} (x-y),
 \label{eq:ymcal}
\end{eqnarray}

so the gauge parameter effectively only changes the conformal class of the
metric.

The end result of the calculation is that

\begin{equation}
\int \int dx \: dy L_{3} \star L_{3} = \int dx L_{4} .  \label{eq:ymres}
\end{equation}

It would be very interesting if on could interpret the symmetric field $G^{\nu
\rho}$ appearing in (\ref{eq:ymstr}) as the summetric tensor field that appears
in string theory (since there is not any natural constant appearing in the
transformation laws, except perhaps the zero mode of L). An interesting
interpretation goes as follows: By equations
(\ref{eq:winfem}),(\ref{eq:winfym}) we see that the parameters for the w gauge
transformations are the tensor fields over the group manifold. If we interpret
the group manifold as arising from compactification and the particle theory as
arising from string theory then by the duality between gauge parameters and
fields (as in (\ref{eq:dual})) the above interpretation is reasonable. We can
look for dimensions of the fields appearing in (\ref{eq:winfem}). In D
spacetime dimensions the following is true:

\begin{equation}
[A]=\frac{D-2}{2}, \: \:\;\;\;  [L_{{k}}] \equiv [L_{\nu _{1} \ldots \nu
_{k}}]=\frac{D(1-k)+2k-4}{2} ,
\end{equation}
\begin{equation}
[L_{{k_{1}}} \otimes L_{{k_{2}}}]=[L_{{k_{1}+k_{2}}}] \Longrightarrow D=4 .
\label{eq:D4again}
\end{equation}

So again we see that $\dot{w}_{\infty}$ leads naturaly to four dimensions.
Now let us consider the background field method, in view to possible
applications to the background independence problem. In view of the previous
considerations we expect that the background breaks the symmetry. This is easy
to see in the YM context too: Consider the background symmetry: At the first
level (which is the usual gauge transformations) we know that
\begin{eqnarray}
A_{\mu}=A_{\mu}^{B}+A_{\mu}^{Q},\\
\delta_{B} A_{\mu}^{B} =[D_{\mu}(A^{B}),\Lambda ], \\
\delta_{B} A_{\mu}^{Q} =[A_{\mu}^{Q},\Lambda ],
\end{eqnarray}

and the usual algebra is obeyed. This feature cannot be maintained at the
higher levels:

\[\delta_{k}A \neq \delta_{k}^{B}(A^{B})+\delta_{k}^{B}(A^{Q}).\]

Therefore we see that the background breaks again the $\dot{w}_{\infty}$
symmetry.

In this section we saw that the $\dot{w}_{\infty}$ symmetry can be realised in
the case of gauge fields in a naively trivial way (modulo global topological
considerations), namely by field dependent redefinitions of the gauge
parameters. In the same spirit it can be realised in the string field theory
level, as we will see in the following section. Also field dependent
redefinitions of the gauge parameters in the sigma model generate gauge
transformations of the target space fields:

Consider the vertex insertion (in the loop variable approach \cite{kn:sat})

\[A_{\mu} \partial _{+} X^{\mu} +E_{\mu \nu} \partial _{+} X^{\mu} \partial
_{-} X^{\nu} + \partial _{+} L . \]
By partially integrating the last term it is well known and obvious that we
obtain the gauge variation law for the U(1) field. If we consider the string
field dependent redefinition of the gauge parameter L as follows
\[ L'=L_{\nu} \partial_{-} X^{\nu}, \]
we see that we obtain the gauge transformation laws for the symmetric and
antisymetric part of $E$, with equal gauge parameters.

\section{String Field Theory/Background independence}
Witten's open string field theory, and Zwiebach's closed string field theory
are both foliation gauge fixed, in the sense that at each vertex of the
critical graph three critical edges emerge. This means that they use a prefered
covering of the moduli space (actually a prefered class of coverings connected
by renormalisation group equation \cite{kn:hat}). However this must pressumably
be a gauge fixed picture of a theory that does not care about which foliation
picture we use. So our known string field theories must be supplemented with a
ghost action.
We will now prove that $\dot{w}_{\infty}$ is also a residual symmetry of
string field theory written around a trivial sigma model background. Usually
one defines the string products by fixing a metric on the world sheet, say
midpoint interaction, or minimal area. The way we will be thinking about this
is that it actually corresponds to a partial gauge fixing of the string field
theory Lagrangian, with residual symmetry the conformal one at the first
quantized level , and $\dot{w}_{\infty}$ at the second quantized level, which
is basically a symmetry of possible field redefinitions, so again the orbits of
this symmetry are the possible distinct fields (vacua) as we also saw in the
first quantized level (\ref{eq:twispace}). If we do not take this partially
gauge fixed theory, we must include all metrics, and all the foliations on the
world sheet. The non-partially gauge fixed theory has actually bigger symmetry,
which is not known at this point. The merit of this might be in due time the
solution of backgroun
d independence of string field theory, the problem in the present formulation
being essentially that we do not use the $\dot{w}_{\infty}$ ghosts, which in
turn somehow enforces field equations for the background. The idea is that
putting the string to a background corresponds to puncturing the world sheet
with background current insertions, therefore to spliting the naive foliations
of the world sheet. But this is exactly what $\dot{w}_{\infty}$ does, so in
particular the $w_{m}$ generator maps $\cal H$ to $\otimes_{m}\cal H $, where
$\cal H$ is the string hilbert space. This is exactly in accordance to what we
saw in the YM case.
Let me now be explicit and consider Witten's open string field theory.
Consider the following thansformation rules:

\begin{equation}
\delta _{k+1} A=\frac{1}{k+1} (Q+A) (A^{k} \Lambda _{k+1}+(-1)^{\Lambda}
A^{k-1} \star \Lambda _{k+1} \star A +more) ,\label{eq:winfsft}
\end{equation}

where Q+A acts in the adjoint representation with respect to the
$\star$-product.

One can again verify that the same algebra holds:

\[[\delta_{k},\delta_{l}] A^{m} = \sum_{i=1}^{k+l-1} \delta_{k+l-i} A^{m} \]

Where we have to allow all possibilities allowed by group theory. As an example
to illustrate what I mean let's take $\Lambda_{2}$ with ghost number -3 and
grassmann odd coefficient functions. Then $\delta_{2}A$ has ghost number -1 and
corresponds to the $F \star F$ theory of appendix \ref{ap:coo}. However a
second quantised string state of ghost number -1 can be assosiated to the half
of the string, by noticing that -1+3/2-1=-1/2. So $w_{2}$ changes half of the
string in a way that it is a symmetry. Changing half of the string can be
thought of as inserting a vertex operator in the middle of the string, which
will change the foliation pattern around it and will introduce critical points
and/or branch cuts. Isolated critical points will be assosiated to closed
string background inserted in the middle of the string. Similarly for the
higher order generators. This is of cource a symmetry of Witten's action CS(2).

Let us consider
\[ \delta _{1} A=Q \Lambda _{1} +A \star \Lambda _{1} - \Lambda _{1} A, \]
\[ 2 \delta _{2} A= QA \star \Lambda_{2}-A Q \Lambda _{2} -Q \Lambda _{2} A +
\Lambda _{2} QA +A^{2}\Lambda_{2}-\Lambda_{2} A^{2},\]
Then one can show that
\[[\delta_{1}(\Lambda_{1}),\delta_{2}(\Lambda_{2})]=\delta_{2}(\Lambda_{1}
\star \Lambda_{2}-\Lambda_{2} \star
\Lambda_{1})+\delta_{1}(-\frac{1}{2}(\Lambda_{2} Q \Lambda_{1} +\Lambda_{1} Q
\Lambda_{2}).\]

So we see that we obtain an algebra which is $\dot{w}_{\infty}$ plus lower
dimensional terms (like in YM). However if $Q\Lambda_{k}=0$ then we obtain the
usual algebra. This looks like pregeometry (\cite{kn:hor})

Let me now discuss non-polynomial closed string field theory. There it is known
\cite{kn:zwi} that a geometric identity as well as a homotopy assosiative
structure exists. It is therefore natural to look for a $\dot{w}_{\infty}$
structure.

We postulate the following classical transformation rule:

\begin{equation}
\delta _{k} (\Lambda _{k} ) \Psi =\sum _{m_{1}, \ldots , m_{k} =0}^{\infty}
 \frac{k^{m_{1}+\ldots +m_{k}}}{m_{1}! \ldots m_{k}!} [\Psi ^{m_{1}}, [ \ldots
,[\Psi ^{m_{k}}, \Lambda _{k}]\ldots]]  .     \label{eq:clowinf}
\end{equation}

In the above equation the braces denote the string field product at zero genus.
The usual gauge transformation is in this notation the $\delta _{1}$.

One can prove by using the geometric identity that the above algebra closes up
to the classical field equation to give again $\dot{w}_{\infty}$. The form of
the geometric identity that is most useful is as follows:

\begin{eqnarray}
\sum _{l,k=0}^{\infty} \frac{1}{l!k!}( {(-)^{\Lambda _{1}} [\Psi ^{l},\Lambda
_{1} ,[\Lambda _{2},\Psi ^{k}]]+(-)^{\Lambda _{2} (\Lambda _{1} +1)} [\Psi
^{l},\Lambda _{2} ,[\Lambda _{1} ,\Psi ^{k}]] +\nonumber \\
+[\Psi ^{l},[\Lambda _{1},\Lambda _{2},\Psi ^{k}]]+(-)^{(\Lambda _{1} +\Lambda
_{2})}[\Psi ^{l},\Lambda _{1},\Lambda _{2},[\Psi ^{k}]])=0   .
\label{eq:geocloid}
\end{eqnarray}

As an illustration one can easily prove that
\begin{equation}
[\delta _{1} (\Lambda _{1}),\delta _{2} (\Lambda _{2})]=(1-2)\delta
_{2}(\Lambda _{2}'=\sum _{n=0}^{\infty} \frac{k^{n+1}}{n!} [\Lambda
_{1},\Lambda _{2},\Psi ^{n}])  ,       \label{eq:clwin}
\end{equation}

plus a zero on-shell term. $\Lambda _{1}$ is odd and $\Lambda _{2}$ is even.

So we conclude that the closed string field theory around the trivial sigma
model background has $\dot{w}_{\infty}$ symmetry. Also if we shift the
background as in (\ref{eq:2}) we cannot maintain the symmetry, as we also found
out in the sigma model version of the theory. The proof is the same as in the
YM case and therefore omitted.

In the beginning of the section we claimed that what is missing from the
present formulation of string field theory are the foliation ghosts, or
equivalently formulation of the theory for any ghost number, and for any first
quantized background. At the BV procedure there appear a lot of ghost fields at
the quantum level, and it would be very interesting if they can be interpreted
as second quantized foliation ghost fields. However most likely this is not
enough and we should really include foliation ghosts at the first quantized
level, in order to make deformations of the complex structure dynamical, e.g
the Liouville mode. The need to include all ghost numbers can be seen in the
topological phase of string theory too: If the topological phase is to be
labeled by the vacuum manifold (or orbifold) of some instanton-like action,
therefore by zero action configurations, then we should be able to write the
action in a ghost number independent form, in the sence that

\begin{equation}
(QA=0) \wedge (L(A)=0)  \Rightarrow Q(b_{0}A)=0.
\end{equation}

Let us consider the action (which is reminiscent to earlier proposals for
gravity actions being summations of topological invariants):

\begin{eqnarray}
L=CS_{2}(A)+CS_{3}(A')+\sum_{k}B_{k} \star b_{0}
(\delta_{k}(\Lambda_{k}=C_{k})A)\nonumber \\
 +\sum_{k}B_{k}' \star (b_{o} \star(\delta_{k}(\Lambda
=C_{k}')A')-(\delta_{k}(\Lambda=C_{k})A))+more , \label{eq:zwiact}
\end{eqnarray}

with classical field equation

\begin{equation}
 F(A)+a(F \star F)(A')=0, \label{eq:64}
\end{equation}

to be compared with (\ref{eq:nest})

Note that in the ghost action coming from $\delta_{2}$ there is a term which
modifies Q to $Q+B\star C$, which might be helful in proving background
independence.

\section{Conclusions/Prospects}
In this paper we saw emerging an interesting new mathematical structure which
we called $\dot{w}_{\infty}$, which is  a twistor inspired modification of the
usual $w_{\infty}$. This symmetry exists both in the first quantized and the
second quantised formalism (at the tree level).

 In the first quantized language it means that Whitehead decomposition of
vertices is a symmetry of the theory, and this implies very strong constraints
for the background fields, in the form of a coproduct structure where high rank
tensor fields appear as (contracted) products of lower rank fields. One can
conjecture that this is true for the full spectrum of the theory (at the
unbroken phase, say constant backgrounds). Verification of this will have to
include loop calculations, because states with less boxes in their Young
tableaus than their mass level will have to involve second or higher order
world sheet derivatives.

 In the second quantized language we made plausible the interpretation that
since $\dot{w}_{\infty}$ corresponds to field dependent redefinitions of the
gauge parameters, it actually is a Fock space generalisation of the (Hilbert
space) notion of gauge symmetry. This means that one or more pure gauge
particles can be emitted from a propagating gauge boson. This point however
needs more clarification. We also showed in the YM case that homotopy
contaction along a purely longitudinal momentum state results to constructing
the four point YM vertex out of two three point vertices. It would be nice if
the same is true if we contract along a longitudinal polarization state. In the
particle case the same mechanism in one space dimension Lorenz covariantises
the Newtonian mechanics action. In the string field theory case a plausible
conjecture would be that this mechanism corresponds to pinching of surfaces
(therefore to factorisation of amplitudes).  We also saw that homotopy
assosiativity is what makes $\dot{w}_{\
infty}$ work.

We also constructed a second quantised Twistor space for string theory as the
space of orbits of the $\dot{w}_{\infty}$ and we compared it with its first
quantised counterpart. It is perhaps an interesting problem to find a
correspondence between the corresponding orbifold points. In the topological
phase we pressumably need an action on this twistor space, with dynamical
w-gravity fields and foliation ghosts. Integration over the foliation
(anti)ghosts will probably turn out to be equivalent to integrating over the
Universal Teichmuller space of the theory. A speculative $\dot{w}_{\infty}$
covariant formulation of string field theory will be very useful for studying
background independence because it will necessarilly include arbitrary
foliations, where arbitrary tensor fields would be inserted (the rank of the
tensor field should of course be equal to the rank of the foliation split).
The fact that the usual formulation of both open and closed string field theory
around the trivial background is $\dot{w}_
{\infty}$ invariant in a trivial way  means that  we are actually integrating
over the Whitehead inequivalent foliations, therefore over all inequivalent
deformations of the complex structure. However for a background independent
formulation we need arbitrary foliations, so the dilaton factor of the metric
enters the scene, and the field content gets extended to twenty seven
dimensions. Also the foliation ghosts must enter the picture to exactly provide
integration over all Whitehead equivalent foliation pictures (therefore over
the dilaton field) and the foliation antighosts over all Whitehead inequivalent
foliation pictures, therefore over the tangent space of the Teichmuller space.
We do not expect the foliation ghosts to provide any extra central charge,
however this is something that needs to be verified explicitly. The above
approach might solve the long standing problem of background independence. Also
as we showed in the sigma model action it provides a natural scenario for
compactifying down to fo
ur dimensions and obtain the standard model symmetry (almost) a la
Kaluza-Klein. We also saw that the tachyon in the spectrum might not be as big
a problem as it is usually thought of (in a theory of both open and closed
strings), because its vanishing is equivalent to the vanishing of closed string
tadpoles.

Another interesting problem is the Hamiltonian realization of
$\dot{w}_{\infty}$, because the twistor notion is basically a Hamiltonian one,
and because in conformal theory language, it is X,$\Pi$ that are canonically
conjugate, not X,$\partial X$. For the one dimensional target space case this
is readily accomplished:

\begin{eqnarray}
\delta _{k} \pi =\pi ^{k} \delta \tau _{(k)} ,\\
\delta _{k} x=-k x \pi ^{k-1}  \delta \tau _{(k)} ,
\end{eqnarray}

in a way that the canonical commutation relations are preserved:

\begin{equation}
\delta _{k} ([x,\pi])=0.
\end{equation}

Note that again this symmetry mixes mass levels and acts like a (momentum
dependent) dilatation on the space coordinate.
Generalizations to higher dimentional spaces and to the string are again less
clear, and under investigation, in the framework provided in \cite{kn:sieax}.

\section{Addendum}
After completion of this work we received some preprints relevant to our
findings. In \cite{kn:boe} it is claimed that W transformations are homotopy
contractions of ordinary gauge transformations. We support this view too, and
moreover we saw that performing the homotopy contraction on lower order
vertices actually covariantizes the action in the case of the particle and YM.
Also in \cite{kn:hat} it is also realized that fields might be missing for a
background independent formulation of string field theory.

\section{Acknowledgements}
I would like to thank S. Garoufalides, J. Kwapisz, M. Sotiropoulos and
especially W. Siegel for discussions and the ITP at Stony Brook for providing a
stimulating environment for self development.

\appendix
\section{Covariant formulation of the sigma model}    \label{ap:cov}
The inclusion of massive modes to the sigma model is possible to be
accomblished in a covariant way provided that we add one more dimension to it
\cite{kn:siepri}. This is the usual trick of obtaining massive representations
by adding one more dimension to the massless ones (e.g \cite{kn:sie}). In order
to have a uniform treatement it is convenient to add one more dimension even to
the massless sector, by simultaneously including a two dimensional dilaton with
central charge $c_{\phi}=-1$:

\begin{equation}
\mu ={m,27} \: \: \: \: m=1 \ldots 26 , \: \: \: c_{\phi}=-1  .
\label{eq:fiecont}
\end{equation}

Let us write down the action which corresponds to the  kinetic term, the
Liouville term and the $\Box \!  \Box \otimes \Box  \! \Box$ closed string
state.

\begin{equation}
L_{kin,Liou}=\partial _{+} X^{\mu} \partial _{-} X^{\nu} G_ {\mu \nu} -
\frac{27-26}{48 \pi} (\frac{1}{2} \partial _{+} \phi \partial _{-} \phi -Q
\hat{R} \phi +\mu _{tot} e^{\alpha \phi}) ,    \label{eq:kili}
\end{equation}

\begin{equation}
L_{mass} = A_{(\mu \nu)(\rho \sigma)} \partial _{+} X^{\mu} \partial _{+}
X^{\nu} \partial _{-} X^{\rho} \partial _{-} X^{\sigma} . \label{eq:mass}
\end{equation}

By choosing
\begin{equation}
Q_{ren}=i \sqrt{\frac{2}{3}}     ,        \label{eq:Q}
\end{equation}

 we get total central charge equal to zero and a negative kinetic term for the
Liouville field which therefore is interpreted as time coordinate
(\cite{kn:pol})

It is clear in principle to go back to the 26 dimensional formalism from this
27+1 dimensional one, by interpreting the $X^{27}$ as a (compactified or not)
Liouville field. This should include a renormalisation of the background charge
$Q$ coming from the massive terms. For example in \ref{eq:mass} we have the
term

\begin{equation}
<A_{(mn)27 \: 27} \partial _{+} X^{m} \partial _{+} X^{n} > (\partial _{-}
\phi)^{2}   .     \label{eq:Qren}
\end{equation}

There is a subtle point that we must set the Liouville momentum equal to the
square root of the mass level, which is more reminiscent of a random walk on a
mass level lattice than compactification a la Kaluza-Klein (\cite{kn:sat}).

\section{Quantisation of the w-particle}

Let me now construct the quantum action. The complete BV quantisation will
hopefully follow in a future publication, here we will only provide a truncated
action which corresponds to a codimension two hypersurface in the BRST extended
phase space.

Consider as gauge fixing condition the following:

\[L_{fix}=\sum_{i=2}^{\infty}(h_{i}(\tau)-1)^{2}.\]

The above choice of gauge fixing term deserves some comments: First of all, in
string theory one always (to my knowledge) uses the maximally weighted fixing
term, so as to eliminate the gauge fields. I do not do that here, to see if
their are any lessons to be learned by not doing so. Second, this gauge fixing
term will lead to convergence of series problems later. It will be seen that
actually is required to take gauge fixing term that drops to zero as $i$ goes
to infinity (and therefore that the usual particle is the singular limit of
this process, $ h_{i}=0$ if i bigger or equal to three). However, due to the
singular character of this limit when background fields exist the usual
problems (of non background independence) arise.

{}From the classical laws one can read off the BRST transformation for the
classical fields:

\[\delta_{\Lambda}h_{i}(\tau)
=[\frac{d}{d \tau}(h_{i}c)-\sum_{k=1}^{i-1} (i-k+1) h_{i-k+1} c_{k}] \Lambda
,\]

\[\delta_{\Lambda}\dot{x}=(\ddot{x} c +\sum_{k=1}^{\infty}\dot{x}^{k} c_{k})
\Lambda .\]

Using the above laws it is easy to write down the reduced ghost action

\[L_{g}=\sum_{i=2}^{\infty} b_{i}(\frac{d}{d \tau}(h_{i}c)-\sum_{k=1}^{i-1}
(i-k+1) h_{i-k+1} c_{k}), \]

the reduction in the BRST phase space being given by the two conditions:

\[b=\frac{\sum b_{i} \dot{h}_{i}}{\sum \dot{h}_{i}}=\frac{\sum b_{i}
h_{i}}{\sum h_{i}} .\]

Relations of this form are to be expected, since the foliation antighosts
integrate over the Teichmuller space of a given equivalence class of foliation
splits, and the union of these spaces is the usual Teichmuller space of the
theory. Therefore integration over b is replaced with integration over the
$b_{k}$'s.

As in the case of Yang-Mills theory, we find the antighost transformation law
as to cancel the BRST transformation of the gauge fixing term, and the ghost
transformation law from the ghost action.

The results are

\[\delta_{\Lambda}b_{i}=2(h_{i}-1) \Lambda, \]
\[\delta_{\Lambda}c=\dot{c} c \Lambda ,\]

\[\delta_{\Lambda}c_{k}=(\frac{\sum_{\tilde{i}=2}^{\infty}
\tilde{i}(\sum_{l'=k}^{\tilde{i} +k-2} (\tilde{i} +k- l')h_{\tilde{i}+k- l'}
c_{l' -k+1})}{\sum_{\tilde{i}} \tilde{i} h_{\tilde{i}}}c_{k}+\dot{c_{k}} c)
\Lambda \]

I have not yet checked explicitly if this is nilpotent on the $c_{k}$, but if
the above mentioned gauge fixing is allowed and complete it must be so, up to a
total derivative, because the gauge fixed action is BRST invariant.

Let us now be more rigorous and consider a simplified (truncated) model with
action
\begin{equation}
S_{2}=\int h_{2} \dot{x} ^{2} + (h_{2}-1)^{2}+b_{2}[\frac{d}{d \tau} (h_{2}
c)-2 h_{2} c_{1}] ,   \label{eq:simple}
\end{equation}

and truncated nilpotent BRST
\begin{eqnarray}
\delta_{\Lambda} h_{2}=(\frac{d}{d \tau} (h_{2} c) -2 h_{2} c_{1}) \Lambda
\nonumber ,\\
\delta_{\Lambda} \dot{x} =(\ddot{x} c +\dot{x} c_{1}) \Lambda ,  \nonumber \\
\delta_{\Lambda} c_{1} =\dot{c}_{1} c \Lambda , \\   \label{eq:restrbrs}
\delta_{\Lambda} b_{2} =2 (h_{2} -1) \Lambda , \nonumber  \\
\delta_{\Lambda} c =\dot{c} c \Lambda ,\nonumber
\end{eqnarray}

and the path integral (note that in the Lagrangian formalism we do not
integrate over c)

\begin{equation}
A=\int dc_{1} db_{2} dh_{2} dx e^{-S_{2}}   .    \label{eq:pathint}
\end{equation}

This integral can be easily performed to give

\begin{equation}
A= \int dx e^{-S_{eff}}   ,\label{eq:seff}
\end{equation}

with

\begin{equation}
S_{eff}= -ln(\sqrt{\pi} /2)+\frac{3}{2} \dot{x}^{2} -\frac{1}{8} \dot{x}^{4}
+\frac{1}{24} \dot{x}^{6} +\frac{1}{64} \dot{x}^{8} +more . \label{eq:eqeff}
\end{equation}

We see that integration over the unphysical fields gives higher order terms in
the lagrangian. We conjecture that by performing the all loop calculation we
will end up with the Einstein action, or in other words that Newtonian physics
is homotopy contraction of Lorenz physics. We will not perform an explicit
calculation here, but in  section three  we saw an indirect way to prove this.

Now consider the coupling to W-gravity target space currents $ f_{i}(x(\tau))$
with action
\[S=\int_{0}^{1} d \tau \sum_{i=2}^{\infty} h_{i}(\tau) \dot{x}(\tau)^{i}
f_{i}(x(\tau)).\]

These currents of course play the role of background fields and of generalised
coupling constants, as in the usual two dimensional non linear sigma model. In
order to preserve the above symmetry, we actually cannot have arbitrary
background fields. In this one dimensional target space toy model the
consistency condition is actually very strong, namely that all the coupling
functionals are equal up to proportionality constant. Also one can contemplate
about the case that the usual transformation laws for the gauge fields are
obtained modulo a (field dependent) diffeomorphism, a situation that would be
reminiscent to the homotopy assosiative algebra of states of string field
theory.

Now let's consider the case of higher dimensional target space:
\[S=\int d \tau \sum_{i=2}^{\infty} h_{i}(\tau) \dot{x}^{\mu_{1}} \ldots
\dot{x}^{\mu _{i}} g^{(i)}_{\mu _{1} \ldots \mu_{i}}(x(\tau)).\]

The law

\[\delta _{k} \dot{x} ^{\mu} = g^{\mu}_{\mu_{1} \ldots \mu_{k}} \dot{x} ^{\mu
_{1}} \cdots \dot{x}^{\mu _{k}} \delta \tau _{k}, \]

with the x inert if the background fields are x-dependent (so in this approach
w gravity is broken) and the h's transforming as before is a symmetry provided
that the following consistency condition holds for each n,k (symmetrisation of
indices is understood throughout):

\[g_{\mu \mu_{k+1} \cdots \mu_{n}} g^{\mu}_{\mu_{1} \cdots \mu_{k}}=g_{\mu _{1}
\cdots \mu _{k} \cdots \mu _{n}, \]

so in effect, only the first two levels of fields are independent. (One can
postulate appropriate transformation laws for the x's such as to retain the w
algebra, but then the action is invariant again modulo x dependent
diffeomorphisms).

Now let me start explaining the background independence of the theory. Suppose
we have been able to construct a nilpotent BRST operator for our model. This
can be read from the above expressions, but I haven't yet checked the
nilpotentcy on the ghosts.
Then, using the usual tricks of BRST, we expand our field in the superspace
basis provided by the ghost coordinates:

\[\Phi =\phi (x)+\phi ^{\ast}(x)c_{N}+ c \psi (x) +c c_{N} \psi^{\ast}(x)+ \sum
(c_{k} \psi _{k} (x)+\hat{c_{k}} \psi ^{\ast} (x)) +\]
\[\sum (cc_{k} \psi _{k}'+c \hat{c_{k}} \psi_{k}'^{\ast}) + \sum (c_{k} c_{l}
\psi _{kl}+\hat{c_{k}} \hat{c_{l}} \psi_{kl}^{\ast}) + infite \: more ,\]

where in the above formula  \[\hat{c_{k}} \equiv \prod _{l \neq k} c_{l} =c_{1}
\ldots c_{k-1} c_{k+1} \ldots,\]
and also \[c_{N} \equiv \prod _{l=1}^{\infty} c_{l}. \]

If $c_{k}$ represents a state filled with a k-ghost, then $\hat{c_{k}}$
represents a k-antighost state. Both these states are grassmann odd. Note the
proliferation of ghosts, which are needed to gauge away `fake' Hilbert spaces,
a phenomenon that will be even more striking in the YM and the BRST string
field theory case.

The problem with the BRST operator is that one cannot explicitly invert the
momenta, to get the velocity in terms of them. (A better approach might have
been to use hamiltonian BRST from the beggining). Suppose that after field
redefinitions for the ghosts the expression for the first quantised sigma model
Q  is:

\[ Q=c\sum_{i} f_{i}' p^{i} + \sum_{k} c_{k} p^{k} f_{k} +infinite \: more. \]

One can  calculate the expression

\[ S=\int dx dM \Phi Q \Phi \: \: \: \: dM = \ldots dc_{2} dc_{1} dc ,\]
to find the second quantised action, if one wishes. It would be interesting to
see if the higher derivative couplings to the ghosts somehow provides a natural
intrinsic mechanism for regulating the theory.

\section{Open string theory with closed string tadpoles} \label{ap:coo}

As we saw it will be beneficial to construct theories written on arbitrary
ghost number fields. In the fermionic string case it is known that there exists
a picture changing operator. However let us consider a toy bosonic model., with
interpretation of describing open strings in a closed string tadpole
background, or equivalently gauge fixed in a target space tachyonic way
(figs.4,9).
Let me consider a 2-nd quantized string field functional A(X($\sigma$),$\tau$)
with ghost number G. Then since Q increases ghost number by 1 we deduce that
the gauge parameters have ghost number G-1, and so for closure the $*$
operation has ghost number 1-G (see appendix B). Consider the following general
expression:
\begin{equation}
A*\cdots*A*QA*\cdots*QA,
\end{equation}
where Q appears q times and A p+q times. A simple counting gives us the ghost
number of this expression:\[G+2p+q-1.\]
So for a theory with fixed G,we can consider terms of the above form, for
\begin{equation}
G+2p+q-1+G^{0}=0,
\end{equation}
where $G^{0}$ is the ghost number violation of the integration sign.
Let me consider a naive example of such a theory. A possible  interpretation of
the contractible loop foliations on the open string as the result of the
insertion of a background closed string (closed string tadpole).  Take G=-1 and
$G^{0}$=-3. Then it is trivial to check that the possible terms appearing are:
\[A*QA*QA,A*A*A*QA,A*A*A*A*A.\]

Consider the following 2-nd quantized lagrangian density:

\begin{equation}
L=aA*QA*QA+bA*A*A*QA+cA*A*A*A*A.
\end{equation}
By demanding invariance of the action under

\begin{equation}
\delta(A)=Q\epsilon+A*\epsilon-\epsilon*A,
\end{equation}
it is straightforward to see that
\[3a=2b=5c,\]
and that the E-L equations of motion are
\begin{equation}
F*F=0,
\end{equation}
where as usual (\cite{kn:witb})
\begin{equation}
F=QA+A*A,
\end{equation}
which means that the second Chern class of the line bundle of string
functionals is trivial. Actually, since F, $*$, $\int$ are the only gauge
covariant quantities in all these theories, the equations of motion one gets
have to be the vanishing of some topological Chern-like class of the bundle.
Let me now explain the generic case: Suppose I want to write down a theory with
high order interaction term, corresponding to many (three or more) strings
meeting at the interaction vertex (figure 6). By performing a Whitehead
decomposition I can reduce the interaction to say n minimal interaction points.
Then the number of edges of this reduced diagram (which is equal to the number
of interactions) is 2n+1 which corresponds to a pure potential term
\[\int\underbrace{A*\cdots*A}_{2n+1}.\]

This theory is meant to be equivalent to a theory with minimal vertex of string
states of ghost number G.
Since the $\int$-operation is meant to cancel the total ghost number violation
as a matter of counting of free strings colliding at a vertex, (and $*$ takes
care of the ghost number violation due to the ghost current anomaly on a curved
world sheet) then we can postulate
\begin{equation}
G^{0}=3G.
\end{equation}
Then counting ghost numbers we have
\begin{equation}
0=3G+G(2n+1)+2n(1-G)\Longrightarrow2G+n=0,
\end{equation}
so $G=-n/2$ is always a negative integer or half integer (so a multiple of
-1/2). This is a very strong result corresponding to the fact that actually
changing the ghost number by an additive constant while keeping Q the same is
not permissible (as can be also seen by the fact that both Q,J are part of an
OSP algebra) and the most we can do is just take tensor products of our theory
with itself (also because $Qb\neq 0$).
Also note that G=0 implies n=0, so CS string field theory of the type we
consider  cannot exist for zero ghost number.

By varying the action we only get the term
\begin{equation}
\int\delta A*\underbrace{F*\cdots*F}_{n},
\end{equation}
due to the uniqueness of the ghost number assignement of the integration sign.
The ghost number counting is still checking out fine:
\[0=3G+G+n(G+1)+n(1-G)\Longrightarrow2G+n=0\]
Also note that in all these theories
\[vertices-edges+strings=1\]
so the number of strings that collide on a vertex is n+2.

\end{document}